\documentclass{article}
\usepackage{ipe}

\catcode`@=11
\def\@bindentby#1{%
\begingroup\setbox3=\hbox{#1}\par\noindent%
\dimen2=\linewidth\dimen4=\@totalleftmargin%
\copy3\unskip\advance\linewidth-\wd3\advance\@totalleftmargin\wd3%
\parshape=2\dimen4\dimen2\@totalleftmargin\linewidth%
\everypar{\parshape=1\@totalleftmargin\linewidth}%
\ignorespaces}

\def\@eindentby{
        \advance\linewidth\wd3
        \advance\@totalleftmargin-\wd3
        \everypar{\parshape=1\@totalleftmargin\linewidth}
        \par\endgroup}

\newenvironment{indentby}[1]{\@bindentby{#1}}{\@eindentby}
\catcode`@=12

\newenvironment{algorithm}[1]{\Singlespacing\noindent\ \\
        \mbox{{\bf Algorithm {#1}}} \\[-0.1in]}{\vspace{-0.1in}\mbox{{\bf End of the Algorithm}}\medskip}
\newcommand{\alginput}[1]
{\begin{indentby}{\protect\makebox[0.125in]{}{\em Input:\ \ \ }}#1 \end{indentby}}
\newcommand{\algoutput}[1]
{\begin{indentby}{\protect\makebox[0.125in]{}{\em Output: }}#1 \end{indentby} \vspace{-0.1in}}

\newtheorem{theorem}{Theorem}[section]

\newtheorem{lemma}[theorem]{Lemma}
\newtheorem{corollary}[theorem]{Corollary}
\newenvironment{proof}{\noindent {\bf Proof\,\ }}{\hfill\mbox{\ $\diamondsuit$}
        \bigskip}

\newcommand{\Singlespacing} {\baselineskip 12pt}

\begin{document}
\title{Convex Tours of Bounded Curvature\footnotemark[1]}
\author{Jean-Daniel~Boissonnat\footnotemark[2]
\and    Jurek~Czyzowicz\footnotemark[3]
\and    Olivier~Devillers\footnotemark[2]
\and    Jean-Marc~Robert\footnotemark[4]
\and    Mariette~Yvinec\footnotemark[5]}
\maketitle

\footnotetext[1]{ {\em This work has been supported in part by the ESPRIT Basic
Research Actions Nr. 7141 (ALCOM II) and Nr. 6546 (PROMotion), NSERC, FCAR and F
ODAR.} }

\footnotetext[2]{ INRIA, 2004 Route des Lucioles, B.P.109, 06561 Valbonne
cedex, France\\ Phone~: +33 93 65 77 38, E-mails~:\ firstname.name@sophia.inria.
fr}

\footnotetext[3]{D\'ep. d'informatique, Universit\'e du Qu\'ebec \`a Hull}

\footnotetext[4]{D\'ep. d'informatique et de math\'ematique,
Universit\'e du Qu\'ebec \`a Chicoutimi}

\footnotetext[5]{INRIA and CNRS, URA 1376, Lab. I3S, 250 rue Albert Einstein,
Sophia Antipolis, 06560 Valbonne, France}

\begin{abstract}
We consider the motion planning problem for a point constrained to move along
a smooth closed convex path of bounded curvature. The workspace of the moving
point is bounded by a convex polygon with $m$ vertices, containing an obstacle
in a form of a simple polygon with $n$ vertices.  We present an $O(m+n)$ time
algorithm finding the path, going around the obstacle, whose curvature is the
smallest possible.
\end{abstract}

\section{Introduction}
Consider the problem of moving a point robot
in the interior of a convex polygon containing 
a single obstacle. We are looking for a smooth, closed, convex,
curvature-constrained
path of the point around the obstacle. No source or target position of the
point are specified.

The problem of planning the motion of a robot subject to kinematic constraints
has been studied in numerous papers in the last decade
(cf.  \cite{l-rmp-91}, \cite{ss-ampr-90}).
For example, Reif and Sharir \cite{rs-mppmo-94} studied the problem of
planning the motion of
a robot with a velocity bound amidst moving obstacles in two and
three-dimensional space.  {\'O}'D{\'u}nlaing \cite{o-mpic-87} presented
an exact algorithm solving the one-dimensional kinodynamic motion planning
problem whereas Canny, Donald, Reif and Xavier \cite{dxcr-kmp-93}
gave the first approximation algorithm solving the two and three-dimensional
kinodynamic motion planning problem for a point amidst polyhedral obstacles.

Another aspect of the motion planning problem in the plane
consists in finding paths under curvature constraints.
Dubins \cite{d-cmlca-57} characterized shortest curvature constrained
paths in the Euclidean plane without any obstacle. More recently,
Fortune and Wilfong \cite{fw-pcm-91} gave a decision procedure to
verify if the source and target placement of a point robot may be joined by a
curvature constrained path avoiding the polygonal obstacles.
Their procedure has time and space complexity \(2^{O(poly(n,m))}\), where
$n$ is the number of obstacle vertices, and $m$ is the number of
bits required to specify the positions of these vertices.
Jacobs and Canny \cite{jc-pspmr-89} gave an algorithm
computing an approximate curvature constrained path, and Wilfong \cite{w-mpav-88}
designed an exact algorithm for the case where the curvature constrained 
path is limited to some fixed straight ``lanes'' and circular arc turns between the
lanes. Finally, \v{S}vestka {\em et al.}
\cite{os-plamp-94,kslo-prpph-96} applied the random approach
introduced by Overmars \cite{o-ramp-92} to compute curvature constrained
paths for car-like robots.

Besides heuristic and approximating approaches, an exact algorithmic solution
seems to be difficult to find for the general case. An interesting
direction of research is to design exact algorithms for some
variants of the problem. In this paper, we give an efficient solution for
the problem of computing a smooth closed convex path going
around a single polygonal obstacle
with $n$ vertices inside a convex polygon with $m$ vertices.
We design an $O(n+m)$ time and space algorithm finding a path
of smallest curvature. The idea of the algorithm is to 
compute the curvature constraints
imposed by the vertices of the obstacle. The maximal such constraint 
is then used to compute the smooth closed convex path which must surround the
entire obstacle.

Finally, some extensions of this solution for the
case of numerous obstacles, and for the case of
obstacles coming as queries in a dynamic setting are also presented.

\section{Preliminaries}
Let $E \subset {\rm I\hskip -0.2em R}^2$ be a convex polygon with $m$ vertices
and let $I \subset E$ be a simple polygon with $n$ vertices. The region
$E \setminus int(I)$ represents the {\em workspace} $W$ in which the point
robot can move. A function $p:[0,L] \rightarrow W$ is a {\em smooth path} if
$p(r) = (x_{p}(r), y_{p}(r))$ and the functions
$x_{p}, y_{p}: [0, L] \rightarrow {\rm I\hskip -0.2em R}$
are continuous with continuous derivative (i.e. $x_p$ and $y_p$ must be
in $C^1$). A smooth path $p$ is $closed$ if \(p(0) = p(L)\) and
its right derivative at point 0 is equal to its left derivative at point $L$.
As any smooth path has finite length, we assume
that $p$ is parameterized by arc length.
Such a parametrization is called a {\em normal} parametrization of $p$.
Let $\Theta_{p}(r)$ be the angle made by the tangent of the path $p$
at the point $p(r)$ with the $x$-axis. The {\em curvature} of
$p$ at a point $r$ can be defined as
$\lim_{r' \rightarrow r} \frac{|\Theta_p(r)-\Theta_p(r')|}{|r-r'|}$.
It might be possible that the curvature of a path is not defined at
some points. For example, consider a circular arc extending a line segment
in such a way that the circle containing the arc is tangent to the
line containing the line segment. The curvature of the tangent point
joining the arc and the line segment is not defined. In such a case,
we have to consider the {\em average curvature} of the path.
A path $p$ has its average curvature bounded by some constant $\kappa$ if
$|\Theta_{p}(r_{2}) -\Theta_{p}(r_{1})| \leq \kappa |r_{2} - r_{1}|$,
for all $ r_{1}, r_{2}$. If $\kappa$ is the best bound
possible, we would say that the average curvature of
$p$ is in fact equal to $\kappa$.
Hence, the term curvature used in this paper refers
to the notion of average curvature.
By using this definition, the curvature of a
circular arc of radius $r$ is $1/r$ and the curvature of a line is $0$.

A curvature bounded smooth closed convex path $p$ is a $tour$ of $I$ in $E$ 
if the bounded region of $E$, delimited
by the Jordan curve $p$, is convex and contains $I$.
Note that the points of boundaries of $E$ and $I$ are allowed to
lie on the tour. Finally,
a tour is $optimal$ if its curvature is the smallest possible.

The main problem considered in this paper can be formulated as follows.
Find an optimal tour of $I$ in $E$. We first consider the degenerated case
where the internal polygon $I$ is a single point.

\begin{lemma}\label{lemme1}
For a given convex polygon $E$ and a point $v$ inside $E$, let $C$ denote
a circle of radius $r$ inscribed in $E$, passing through $v$,
and tangent to the boundary of $E$ in two points $p_{1}$ and $p_{2}$.
If the arc $\alpha=p_{1}vp_{2}$ of $C$ is not greater than a semicircle,
then the curvature of any tour of $v$ in $E$
is at least equal to $1/r$.
\end{lemma}

\begin{figure}
\begin{center}
{\def\IPEfile{Fig1.ipe}\begingroup
  \catcode`\%=9\catcode`\!=0\catcode`\-=11\input{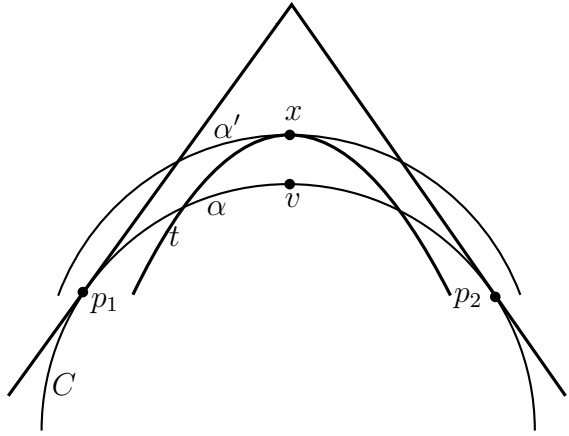}}
\end{center}
\vspace{-3cm}
\caption{\label{fig-lemme1}An optimal tour of a point.}
\end{figure}

\begin{proof}
Let $t$ denote a tour of $v$ in $E$. Such a smooth path must intersect $\alpha$.
Translate $\alpha$ along the bisector of the angle defined by the tangents
of $C$ at $p_1$ and $p_2$. Now, let $\alpha'$ denote the furthest position
of $\alpha$ tangent to $t$ and let $x$ be some tangent point (see
Fig.~\ref{fig-lemme1}).

Suppose that $t$ and $\alpha'$ coincide on a small interval around $x$.
In this interval, the curvature of $t$ is the same as the curvature of
$\alpha'$ which is $1/r$. Now, suppose that $t$ is strictly
below $\alpha'$ just after $x$. Notice that such a tangent point always
exists if $\alpha'$ does not coincide with $\alpha$. It follows from
Lemma~\ref{L:appendix-curv} of the appendix that the
curvature of $t$ is strictly greater than the curvature of
$\alpha'$ which is $1/r$.
\end{proof}

Following this lemma, a circle $C$ inscribed in the polygon $E$ and
tangent to the points $p_1$ and $p_2$ is the {\em critical circle} of
a point $v$ in $E$, if the arc $p_{1}v p_{2}$ of $C$ is
not greater than a semicircle. The arc $p_{1}v p_{2}$ is called the
{\em critical arc} of $v$ in $E$.
Notice that only points lying outside a largest inscribed circle
in $E$ admit critical arcs.

\section{Computing tours}

\subsection{The Case of Given Curvature}
Consider the problem of computing, if one exists, a tour of $I$ in $E$ with
curvature bounded by some given constant $\kappa$.
We present in this section an algorithm solving this problem
in $O(m + n)$ time. The algorithm proceeds by computing a maximal
path in $E$ with curvature bounded by $\kappa$. Note that the value
of $\kappa$ should not be greater than $1/r^*$ where $r^*$
represents the radius of the largest inscribed circle in $E$. This
follows from the fact that any smooth closed convex path of curvature
$\kappa$ should enclose a circle with radius $1/\kappa$ (see
Lemma~\ref{L:appendix-conv}).

\begin{figure}
\begin{center}
{\def\IPEfile{Fig2.ipe}\begingroup
  \catcode`\%=9\catcode`\!=0\catcode`\-=11\input{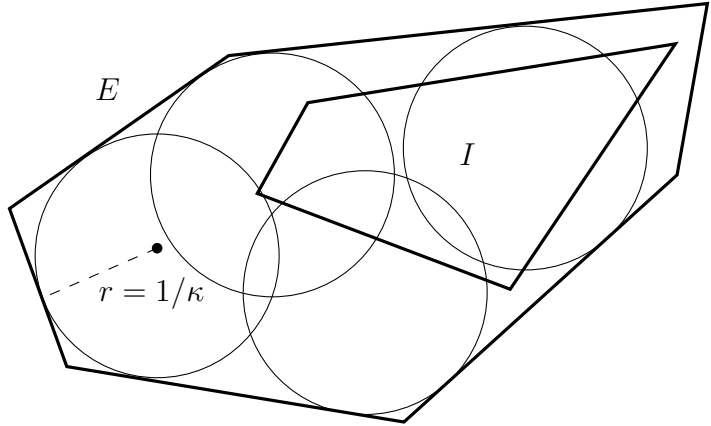}}
\caption{There is no tour of $I$ in $P$ with curvature bounded by $\kappa$.}
\label{fig-max-given-curv}
\end{center}
\end{figure}

Let $S$ be the set
of all circles of radius 1/$\kappa$ inscribed in $E$, and tangent to $E$ in
at least two points. The curve $\zeta$, formed by the boundary of
the convex hull of $S$, is a smooth closed convex path with curvature bounded by
$\kappa$. Such a path $\zeta$ is called a {\em maximal path} in $E$.
It follows from the proof of Lemma~\ref{lemme1} that the convex region
bounded by $\zeta$ contains any smooth closed convex path inside $E$ with
curvature bounded by $\kappa$.
Hence, if $\zeta$ is not a tour of $I$ in $E$,
there exists no tour of $I$ in $E$ with curvature bounded by $\kappa$
(see Fig. \ref{fig-max-given-curv}).

Before we turn our attention to the algorithm verifying the existence of a
tour of given curvature, we introduce some useful concepts. Consider  
the medial axis of $E$ \cite{p-masp-77}. Since $E$ is a convex polygon,
its medial axis corresponds to a tree. Each internal vertex $x$ of this
tree is the center of a circle tangent to three edges of $E$. This circle
is called a {\em Voronoi circle}. We assign to $x$ a weight $w(x)$
corresponding to the radius of its Voronoi circle. Thus, $w(x)$
represents the distance between $x$ and the boundary of $E$.
This weighted tree, rooted at a vertex with the largest weight, 
is called the {\em skeleton tree} and is denoted $SkT(E)$.
It follows from the definition of the medial axis that each edge
of $SkT(E)$ is a straight line segment belonging to the bisector of
some two edges of $E$. It follows also from the definition that each
vertex of $SkT(E)$ has at least two descendants. Finally, we can easily
prove that the weight of any vertex  in $SkT(E)$ is greater than
the weights of its descendants.\footnote{The root may have the same
weight as one of its children if $E$ has two parallel edges.}
This property will be crucial for our algorithms.

We are now ready to present how to compute the maximal path $\zeta$.

\begin{lemma}
\label{lemme2}
Given the skeleton tree $SkT(E)$, the maximal path 
$\zeta$  in $E$ with curvature bounded by $\kappa$ can be computed in
$O(k)$ time, where $k$ is the size complexity of the path.
\end{lemma}

\begin{proof}
Perform a tree traversal on $SkT(E)$.  Each time a vertex $x$ is visited,
such that \(w(parent(x)) \geq 1/\kappa > w(x)\), there exists a
circle of radius
1/$\kappa$ tangent to the boundary of $E$, and centered on the edge joining
$x$ and $parent(x)$. This circle can be computed easily once the
edges of $E$ defining the edge joining $x$ to $parent(x)$ are known.
Then, the subtree of $SkT(E)$ rooted at $x$ is pruned and
the traversal continues from $parent(x)$. In this way,
all the $k$ circles with radius $1/\kappa$ inscribed in $E$ are found in
order of their appearance on $\zeta$. Hence, the maximal path $\zeta$
corresponding to the convex hull of the circles can be obtained easily
by joining two consecutive circles by their common supporting segment.
The $O(k)$ time complexity of the algorithm
follows from the fact that the number of vertices visited during the
transversal of $SkT(E)$ is in $O(k)$.
\end{proof}

\begin{figure}
\begin{center}
{\def\IPEfile{Fig3.ipe}\begingroup
  \catcode`\%=9\catcode`\!=0\catcode`\-=11\input{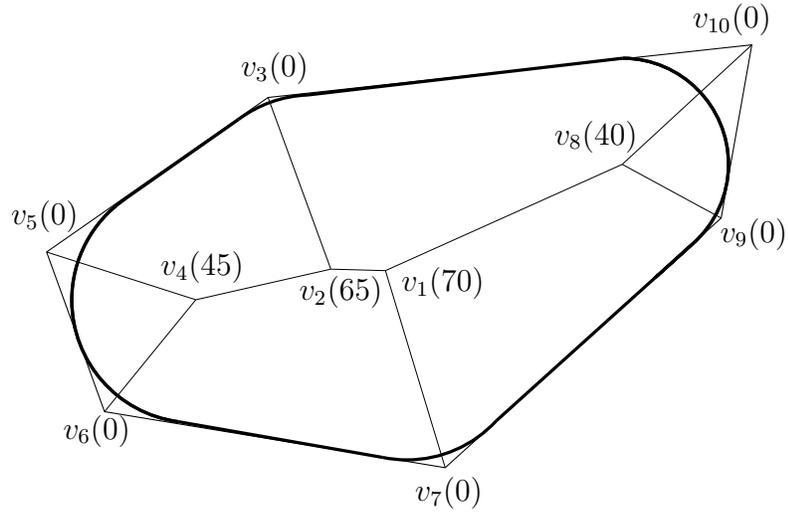}}
\caption{A skeleton tree and a maximal path $\zeta$ of bounded curvature.}
\label{skeleton}
\end{center}
\end{figure}

It should be obvious now how to determine if there exists a tour of
$I$ in $E$ with curvature bounded by $\kappa$. First, compute the medial
axis of $E$ in $O(m)$ time \cite{agss-ltacv-89}. Then,
compute the maximal path $\zeta$ and determine if $I$ lies completely
inside $\zeta$. This latter step can be done easily in $O(n + k)$ time
where $k$ is the complexity of $\zeta$. 
Hence, the algorithm computes, if one exists, a tour of $I$ in $E$
with curvature bounded by $\kappa$ in $O(m + n)$ time.

The notion of maximal path can be related to the notion of
offset curves used in CAD/CAM \cite{Bar92}. The offset curves of convex
polygons can be constructed in linear time without computing the
medial-axis of the polygons \cite{Sut89}.

\subsection{An Algorithm Computing Optimal Tours}
Consider the problem of computing an optimal tour of $I$ in $E$.
An algorithm solving this problem can be sketched as follows.
Find a vertex of $I$ which has the critical arc in $E$ with the minimum
radius. Such a vertex
determines the curvature of an optimal tour. Once the curvature
of the optimal tour is known, a tour can be computed as we described in
the previous section. We present in this section how to implement
this algorithm optimally in $O(m + n)$ time.

\begin{figure}
\begin{center}
{\def\IPEfile{Fig4.ipe}\begingroup
  \catcode`\%=9\catcode`\!=0\catcode`\-=11\input{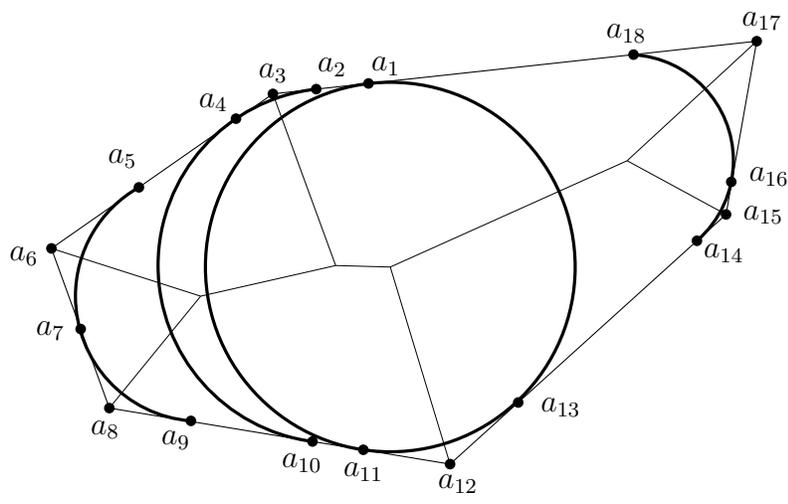}}
\caption{$Arcs(E)$ is determined according to order of arcs' first endpoints.}
\label{arcs}
\end{center}
\end{figure}

We first present the data structures used by the algorithm. Let
$Vertices(I)$ be the list of the vertices of the convex hull of $I$ given
in radial counter clockwise order around the root of $SkT(E)$.
The choice of the root of $SkT(E)$ is arbitrary. We simply
need a point inside a largest inscribed circle in $E$ to simplify the
analysis of the algorithm. This list can be built easily in $O(n)$ time. 
Now, let $Arcs(E)$ be the list of arcs defined as follows.
Consider the Voronoi circles associated with the internal vertices of
$SkT(E)$. The tangent points of these circles with the boundary of $E$
partition each circle into at least three arcs. Each of these
arcs is put in $Arcs(E)$ if it is less than a semicircle.
We also put in $Arcs(E)$ the leaves of $SkT(E)$. These points
represent degenerated arcs. The elements of $Arcs(E)$ must ordered
such that the first endpoints of the arcs appear in counterclockwise
order on the boundary of $E$ (see Fig.~\ref{arcs}).
In the next lemma, we show how to build the list $Arcs(E)$ efficiently.

\begin{lemma}
\label{lemme-arcs}
$Arcs(E)$ can be generated in $O(m)$ time and space.
\end{lemma}

\begin{proof}
Perform a tree traversal on $SkT(E)$.
The traversal can be oriented such that the children of any node
are visited in counterclockwise order.
An arc is produced each
time a vertex $x$ is visited from its parent $v$. 
This arc is less than a semicircle, centered at $v$, and tangent to the two
edges of $E$ whose bisector
contains the edge $vx$ of $SkT(E)$. Finally, a degenerated arc is produced
if $x$ is a leaf of $SkT(E)$.

To see that the arcs are produced in the right order, observe that the
tree traversal can be performed by moving a point $z$ continuously
along the edges of $SkT(E)$. Let $\pi (z)$ be the orthogonal projection
of $z$ on the edge of $E$ belonging to the Voronoi cell on the
right-hand side of $z$ with respect to the direction of the traversal.
Since $SkT(E)$ corresponds to the medial axis of a convex polygon, $\pi (z)$
moves continuously around the boundary of $E$ in counterclockwise
direction. Now, consider the arc computed while $z$ traverses the edge
$vx$ of $SkT(E)$. By construction, the first endpoint of this arc
corresponds to $\pi (z)$ when $z$ coincides with $v$. Thus, the arcs
are produced during the traversal of $SkT(E)$
such that the first endpoints of the arcs appear in
counterclockwise order on the boundary of $E$. 

The $O(m)$ time and space complexities of the algorithm follow from the
fact that $SkT(E)$ has at most $2m-2$ vertices. 
\end{proof}

The points in $Vertices(I)$ and the endpoints of the arcs in $Arcs(E)$
are both sorted according to the radial counterclockwise order around
the root of $SkT(E)$. These two lists will be traversed
simultaneously by the algorithm and the relative order of the
elements of one list with respect to the elements of the other list
is important. Thus, the first element of $Arcs(E)$ should be an arc of a
largest inscribed circle in $E$ and the first element of $Vertices(I)$
should be the vertex just after the first endpoint of the first
element of $Arcs(E)$ in the radial counterclockwise order around
te root of $SkT(E)$.

The variable $V$ will denote  the current element of
$Vertices(I)$ and the variable $A$ will denote the current element of $Arcs(E)$.
We say that the vertex $V$ is {\em before} the arc $A$,
if it precedes the first endpoint of $A$ in the radial counterclockwise
order around the root of $SkT(E)$. $V$ is {\em after} $A$
if it succeeds the second endpoint of $A$ in this order.
For $V$ situated neither before nor after $A$, 
$V$ is {\em inside} $A$ if the ray $pV$ reaches $V$ before crossing $A$,
otherwise $V$ is {\em outside} $A$.

We are now ready to present the algorithm computing an optimal tour
of $I$ in $E$. The aim of the algorithm
is to traverse the list $Vertices(I)$ and localize
each vertex in the planar map generated by the arcs in $Arcs(E)$ and the
boundary of $E$ (see Fig. \ref {arcs}).
Once the cell containing the current vertex is determined,
its critical arc may be computed easily in constant time.

\begin{figure}
\begin{center}
{\def\IPEfile{Fig5.ipe}\begingroup
  \catcode`\%=9\catcode`\!=0\catcode`\-=11\input{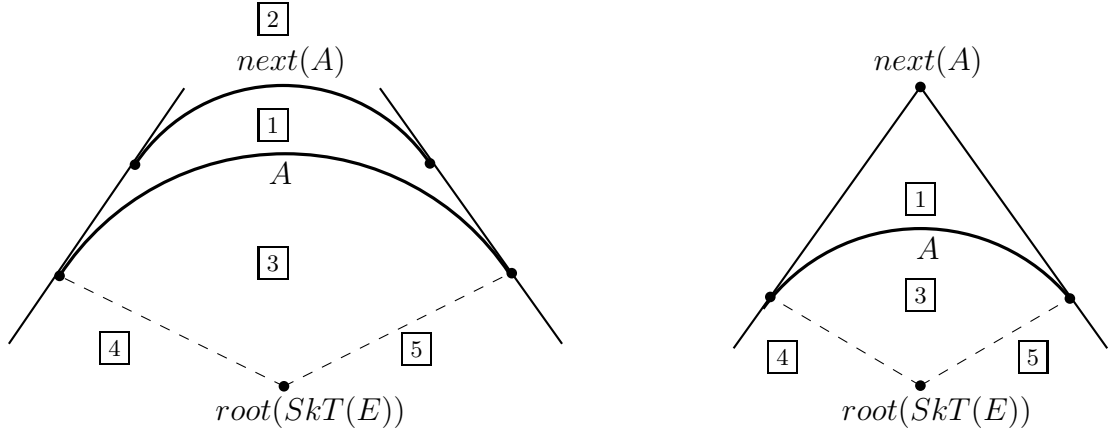}}
\caption{Illustrating algorithm Optimal Tour}
\label{main-algo}
\end{center}
\end{figure}

Each iteration of the main step of the algorithm performs one among five
possible actions. The action depends on the position of $V$ with respect to
five regions determined by the current arc $A$.
Let $next(A)$ denote the successor of $A$ in the list $Arcs(E)$
and let $\overline{next(A)}$ be the smallest arc of the Voronoi circle $C$
extending $next(A)$ and containing all the tangent points between
$C$ and $E$. Notice that $\overline{next(A)}$ lies completely outside $A$.
(see Fig. \ref{main-algo} ). $V$ falls into {\small \fbox{1}},
if it is outside $A$ but not outside $\overline{next(A)}$, and
in Region \fbox{2} if it is outside $\overline{next(A)}$. $V$ is in
Region~{\small \fbox{3}} if it is inside $A$. Finally, $V$ is in
Region~{\small \fbox{4}} if it is after $A$, and in Region~{\small \fbox{5}}
if it is before $A$.

\vbox{
\begin{algorithm}{Optimal Tour}
\alginput{ A convex polygon $E$ of $m$ vertices and a simple polygon $I$ of
   $n$ vertices internal to $E$.}
\algoutput{A tour of $I$ in $E$ with the lowest possible curvature
   bound $\kappa$.}
\begin{enumerate}
  \item Compute $SkT(E)$.
  \item Build the list $Arcs(E)$ sorted by the arcs' first endpoint around 
        the root of $SkT(E)$.
  \item Compute CH($I$) and build the list $Vertices(I)$
        sorted around the root of $SkT(E)$.
  \item $V \leftarrow first(Vertices(I))$. $A \leftarrow first(Arcs(E))$.
  $r \leftarrow radius(A)$.
  \item {\bf while} $Arcs(E)$ is not empty and
                    $Vertices(I)$ is not empty {\bf do}
  \begin{description}
	 \item [\hspace{0.3cm} {\bf case}] the region containing $V$ {\bf do}
	 \begin{itemize}
	   \item[{\small \fbox{1}}]
               $r \leftarrow min(r,{\rm \ radius\ of\ critical\ arc\ of\ }V)$.\\
	       $V \leftarrow next(V)$.
	   \item[{\small \fbox{2}}] $A \leftarrow next(A)$.
	   \item[{\small \fbox{3}}] $V \leftarrow next(V)$.
	   \item[{\small \fbox{4}}] $A \leftarrow next(A)$.
	   \item[{\small \fbox{5}}] $V \leftarrow next(V)$.
     \end{itemize}
  \end{description}
  \item Output the maximal path internal to $E$ with curvature bounded
        by $\kappa=1/r$.
\end{enumerate}
\end{algorithm}}

\subsubsection{The Correctness of the Algorithm}
To prove the correctness of the algorithm, we first have to show
that the algorithm finds the critical arc with the minimum radius.
Thus, by Lemma~\ref{lemme1}, any tour of $I$ in $E$ would have a curvature
at least as great as the curvature of that arc.

The aim of the algorithm is to locate the vertices of $Vertices(I)$ in the
planar map induced by the arcs of $Arcs(E)$ and the boundary of $E$.
A typical cell of that map is bounded by two arcs and by the portions of
two edges of $E$. In
Case {\small \fbox{1}}, the endpoints of $A$ and $\overline{next(A)}$
lie on the same two edges of $E$. This follows
from the fact that the Voronoi circles containing $A$ and $next(A)$ are
centered on the same edge of $SkT(E)$. Hence,
the cell containing $V$ is defined by two edges and two
arcs. The critical arc of any point lying in that cell must be
tangent to the two edges and can be computed in constant time.

In Case {\small \fbox{2}}, the radius of the critical arc of $V$ is
smaller than the radius of the critical arc of any vertex lying in the
cell bounded by $A$ and $\overline{next(A)}$. In Case
{\small \fbox{4}}, neither $V$ nor any subsequent vertex of $Vertices(I)$
will ever lie outside $A$. Hence, the arc $A$ can be discarded in both cases.

Finally, in Cases {\small \fbox{3}} and {\small \fbox{5}},
$V$ lies either inside a largest inscribed circle
or in the cell defined by the arcs $A'$ and $next(A')$,
for some arc $A'$ appearing before $A$ in $Arcs(E)$.
In the former case, $V$ do not admit a critical circle.
In the latter case, since $V$ lies outside $A'$, the arc $A'$
can be discarded only in Case {\small \fbox{2}} of a previous step.
This can only happen if a vertex outside $next(A')$ has been found.
The radius of the critical arc of that vertex is smaller than the
radius of the critical arc of $V$. Thus $V$ can be discarded in both cases.

Hence, the algorithm finds a vertex whose critical arc has the minimum
radius. Then, the maximal path computed in Step~6 must
be a tour of $I$. Otherwise, there would be a vertex of $I$ lying outside
$\zeta$. By construction, the critical arc
of that vertex would have a radius smaller that $r$ which is impossible.

\subsubsection{The Complexity of the Algorithm}
The first two steps of the algorithm rely on well known optimal algorithms.
The convex hull of $I$ can be computed in $O(n)$ time \cite{Gra72}
and the skeleton tree
of $E$ can be computed in $O(m)$ time \cite{agss-ltacv-89}.
In Step~2, the list $Arcs(E)$ can be
built in $O(m)$ time according to Lemma~\ref{lemme-arcs}.
In Step~3, the list $Vertices(I)$ can be built easily in $O(n)$ time.
If the root of $SkT(E)$ lies inside $CH(I)$, $Vertices(I)$ is given
by $CH(I)$. Otherwise, compute the tangents of $CH(I)$ going through
the root of $SkT(E)$ and merge the lower and the upper chains of
$CH(I)$ to produce $Vertices(I)$.
Step~5 represents the core of the algorithm. Each
iteration of the loop takes a constant time. However, as each
iteration removes one vertex of $Vertices(I)$ or one arc
of $Arcs(E)$, the overall time complexity of this step is in $O(n+m)$.
Finally, by Lemma~\ref{lemme2}, the optimal tour $\zeta$ can be constructed in 
$O(k)$ time, where $k \leq m$. Therefore, we obtain the following result.

\begin{theorem}
\label{main-thm}
An optimal tour of a simple polygon with $n$ vertices in a convex polygon
with $m$ vertices can be computed in $O(n+m)$ time and space.
\end{theorem}

The algorithm can be adapted to compute a constrained optimal
tour of $I$ in $E$. Suppose that the tour must to be tangent to
some given lines when passing through some $s$ given points of $E \setminus I$.
Let $E'$ denote the intersection of $E$ with $s$ half-planes
delimited by the given lines, and let $I'$ denote the convex hull of
$I$ and the given $s$ points. Then, the constrained optimal tour is given
by an optimal tour of $I'$ in $E'$.

\begin{corollary}
\label{points-de-passage}
An optimal tour of a simple polygon with $n$ vertices in a convex polygon
with $m$ vertices, constrained to have given tangents when passing
through $s$ given points,
can be computed in $O(n+m+s \log s)$ time and $O(n+m+s)$ space.
\end{corollary}

Finally, we can also consider the problem where the point robot has to go
around many obstacles given as points or polygons lying inside $E$.
In such a case, we simply have to compute the convex
hull of the obstacles and find an optimal tour of the new ``obstacle''.

\begin{corollary}
\label{many-obstacles}
An optimal tour of a set of $n$ points in a convex polygon with $m$ vertices
can be computed in $O(n \log n + m)$ time and space.
\end{corollary}

\section{The Dynamic Setting}
The motion planning problem considered in the previous section can be
reformulated in a dynamic setting. In this case,
we want to preprocess a convex polygon $E$
with $m$ vertices in such a way that for any given query polygon $I$ with $n$
vertices, we can find quickly an optimal tour of $I$ in $E$.

This dynamic problem can be solved by adapting Algorithm Optimal Tour.
In Step~5, if the vertex $V$ lies in Region {\small \fbox{4}}
with respect to the arc
$A$ ,the list $Arcs(E)$ is processed in order but
it is clear that $V$ remains in Region {\small \fbox{4}} with respect to
all other arcs outside $A$. Those arcs correspond to the subtree
of $SkT(E)$ rooted at a
child of the vertex on which $A$ is centered. This subtree
can be skipped in the traversal of $Arcs(E)$.
Hence, the list $Arcs(E)$ is not produced explicitly in Step~2,
but it may be obtained by traversing $SkT(E)$ in Step~5.
The subtree of $SkT(E)$ effectively
traversed is a subset of the subtree of $SkT(E)$ used to generate 
an optimal tour in Step~6.
Thus, the time complexity of Step 5 can be reduced
to $O(n+k)$, where $k$ represents the complexity of the tour.

\begin{theorem}
\label{dynamic-thm}
It is possible to preprocess a convex polygon $E$ with $m$ vertices
in $O(m)$ time and space, so that for any simple polygon $I$
with $n$ vertices, an optimal tour of $I$ in $E$ can
be computed in $O(n + k)$ time, where $k$ is the complexity of the tour.
\end{theorem}

If the obstacle is given as a set of $n$ points instead of a simple
polygon, we simply have to compute the convex hull of these points
and to appply the above result.

\begin{corollary}
\label{point-set-I}
It is possible to preprocess a convex polygon $E$ with $m$ vertices
in $O(m)$ time and space, so that for any set $S$ of $n$ points,
an optimal tour of $S$ in $E$ can be computed in $O(n \log n + k)$ time,
where $k$ is the complexity of the tour.
\end{corollary}

If the curvature of an optimal tour is needed
instead of the tour itself, an alternative solution may be used.
The main problem is still to find a vertex whose critical
circle has the minimum radius. As we saw in the previous section,
this problem can be reduced to a point location
problem in the planar map induced by the arcs of $Arcs(E)$
and the boundary of $E$. For each vertex $v$ of the obstacle, locate
$v$ in the map and compute its critical arc in $E$.

The planar map has $O(m)$ size and it can be decomposed into trapezoids in
$O(m)$ time.  Following the idea of \cite{k-osps-83}, this decomposition can be
preprocessed in $O(m)$ time and space, so that
the point location would be possible in $O(\log m)$ time. Hence, we obtain
the following result.

\begin{theorem}
\label{planar-map-thm}
It is possible to preprocess a convex polygon $E$ with $m$ vertices in
$O(m)$ time and space,
so that for any set $S$ of  $n$ points, the
curvature of an optimal tour of $S$ in $E$ can
be computed in $O(n \log m)$ time.
\end{theorem}

If $m$ is much smaller than $n$,
this method may be interesting even for computing the tour itself.
The following corollary can be used alternatively to
Corollary~\ref{point-set-I}.

\begin{corollary}
\label{point-set-II}
It is possible to preprocess a convex polygon $E$ with $m$ vertices
in $O(m)$ time and space, so that for any set $S$ of $n$ points,
an optimal tour of $S$ in $E$ can be computed in $O(n \log m + k)$ time,
where $k$ is the complexity of the tour.
\end{corollary}

\section{Conclusions}

The paper gives an efficient algorithm computing a smallest curvature motion of
a point robot around an obstacle inside a convex polygon. The solution easily
generalizes on the case of numerous obstacles. We explore the fact
that the resulting path must be convex. In this case, it is sufficient to
compute the curvature constraints imposed by obstacles. The maximal constraint
$\kappa$ is used to compute the maximal curve, internal to the workspace, which must
surround all the obstacles.  The idea works only in the case of convex motion,
and it is not clear how it may be generalized on the case of motion admitting
left and right turns. 

An obvious line of further research is to design algorithms for more general
work\-space.  From the result of \cite{fw-pcm-91} it is possible to draw
a pessimistic inference that a polynomial time algorithm computing
curvature-constrained motion of a point in general workspace may not exist. 
It is natural to ask what are more general settings, that the one studied in
this paper, for which the problem of curvature-constrained motion of a point
admits an efficient solution, and what are the instances of the problem
which are NP-hard.

\appendix
\section{Technical Lemmas}
For completeness, the two technical results on average curvature used
in this paper are presented in this appendix. Their proofs rely
on elementary calculus \cite{Swo91} and differential geometry \cite{Gug63}.

\begin{lemma}\label{L:appendix-curv}
Let $F=(x, f(x))$ be a curve such that $f$ is a convex function in $C^1$.
Let $G=(x, g(x))$ be a curve such that $g$ is a convex function in $C^1$
represented by a circular arc of radius $r$.
If $F$ and $G$ are in contact at the origin (i.e. $f(0)=g(0)=0$ and
$f'(0)=g'(0)=0$) and $F$ lies above $G$ (i.e. $f(x) > g(x)$, for $x > 0$),
the average curvature of $F$ is greater than $1/r$.
\end{lemma}

\begin{proof}
Let $F(t) = (x_F(t), y_F(t))$ be a normal parametrization of $F$ such that
$F(0)=(0,0)$. Let $\Theta_F(t)$ be the angle made by the tangent to $F$ at the
point $(x_F(t), y_F(t))$ with the $x$-axis. The functions $x_F$, $y_F$ and
$\Theta_F$ are related as follows:
$x_F(t) = \int_0^t \cos \Theta_F(u) du$ and
$y_F(t) = \int_0^t \sin \Theta_F(u) du$. Since $F$ is convex and lies above
the $x$-axis for $t > 0$, there is an interval $[0..\epsilon_F]$ on
which $\Theta_F(t)$ is continuous and strictly increasing.
The function $\Theta_G(t)$ is defined similarly and has the same properties.
Hence, there is an interval $(0..\epsilon]$ on which,
either $\Theta_F(t) < \Theta_G(t)$, or $\Theta_F(t) = \Theta_G(t)$,
or $\Theta_F(t) > \Theta_G(t)$.
Suppose that $\Theta_F(t) < \Theta_G(t)$. By definition,
$y_G(\epsilon) > y_F(\epsilon)$ and $x_G(\epsilon) < x_F(\epsilon)$.
Since $x_F$ is continuous, there is a value $\epsilon^*$ such that
$x_G(\epsilon) = x_F(\epsilon^*)$. Furthermore,
$y_F(\epsilon^*) < y_G(\epsilon^*) < y_G(\epsilon)$.
Thus, the point $(x_F(\epsilon^*), y_F(\epsilon^*))$ is below the
point $(x_G(\epsilon), y_G(\epsilon))$ which is impossible.
The case $\Theta_F(t) = \Theta_G(t)$ is even simpler.
Hence, $\Theta_F(t) > \Theta_G(t)$. This implies that
$\frac{\Theta_F(t)-\Theta_F(0)}{t} > \frac{\Theta_G(t)-\Theta_G(0)}{t} = 1/r.$
Thus, the average curvature of $F$ is greater than $1/r$.
\end{proof}

From this technical result, we can obtain the following lemma.

\begin{lemma}\label{L:appendix-conv}
Let $F$ be a closed smooth curve with average curvature $\kappa$.
Then, there exists a circle of radius $1/\kappa$ which lies inside
the convex region delimited by the Jordan curve $F$.
\end{lemma}


\begin{thebibliography}{20}

\bibitem{agss-ltacv-89}
A.~Aggarwal, L.~J. Guibas, J.~Saxe, and P.~W. Shor.
\newblock A linear-time algorithm for computing the {Voronoi} diagram
of a convex polygon.
\newblock {\em Discrete Comput. Geom.}, 4:591--604, 1989.

\bibitem[Bar92]{Bar92}
R.~E.~Barnhill, ed.
\newblock {\em Geometry Processing for Design and Manufacturing}.
\newblock SIAM, Philadelphia, 1992.

\bibitem{dxcr-kmp-93}
B.~R. Donald, P.~Xavier, J.~Canny, and J.~Reif.
\newblock Kinodynamic motion planning.
\newblock {\em J. ACM}, 40(5):1048--1066, November 1993.

\bibitem{d-cmlca-57}
L.~E. Dubins.
\newblock On curves of minimal length with a constraint on average curvature
  and with prescribed initial and terminal positions and tangents.
\newblock {\em Amer. J. Math.}, 79:497--516, 1957.

\bibitem{fw-pcm-91}
S.~Fortune and G.~Wilfong.
\newblock Planning constrained motion.
\newblock {\em Annals of Math. and AI}, 3:21--82, 1991.



\bibitem{Gra72}
R.~L.~Graham.
\newblock A efficient algorithm for computing the convex hull of a
finite planar set.
\newblock {\em Inf. Proc. Letters}, 1:132--133, 1972.

\bibitem{Gug63}
H.~W.~Guggenheimer.
\newblock {\em Differential Geometry}.
\newblock McGraw-Hill, New York, 1963.

\bibitem{jc-pspmr-89}
P.~Jacobs and J.~Canny.
\newblock Planning smooth paths for mobile robots.
\newblock In {\em Proc. IEEE Internat. Conf. Robot. Autom.}, pages 2--7, 1989.

\bibitem{kslo-prpph-96}
L.~E. Kavraki, P.~{\v S}vestka, J.-C. Latombe, and M.~H. Overmars.
\newblock Probabilistic roadmaps for path planning in high dimensional
  configuration spaces.
\newblock {\em IEEE Trans. Robot. Autom.}, 12:566--580, 1996.


\bibitem{k-osps-83}
D.~G. Kirkpatrick.
\newblock Optimal search in planar subdivisions.
\newblock {\em SIAM J. Comput.}, 12:28--35, 1983.

\bibitem{l-rmp-91}
J.-C. Latombe.
\newblock {\em Robot Motion Planning}.
\newblock Kluwer Academic Publishers, Boston, 1991.

\bibitem{o-mpic-87}
C.~{\'O}'D{\'u}nlaing.
\newblock Motion-planning with inertial constraints.
\newblock {\em Algorithmica}, 2:431--475, 1987.

\bibitem{o-ramp-92}
M.~H. Overmars.
\newblock A random approach to motion planning.
\newblock Report RUU-CS-92-32, Dept. Comput. Sci., Univ. Utrecht, Utrecht,
  Netherlands, 1992.

\bibitem{os-plamp-94}
M.~H. Overmars and P.~{\v S}vestka.
\newblock A probabilistic learning approach to motion planning.
\newblock In {\em Algorithmic Foundations of Robotics}, Wellesley, MA, 1995. A.
  K. Peters.

\bibitem{p-masp-77}
F.~P. Preparata.
\newblock The medial axis of a simple polygon.
\newblock In {\em Proc. 6th Internat. Sympos. Math. Found. Comput. Sci.},
  volume~53 of {\em Lecture Notes in Computer Science}, pages 443--450.
  Springer-Verlag, 1977.

\bibitem{rs-mppmo-94}
J.~H. Reif and M.~Sharir.
\newblock Motion planning in the presence of moving obstacles.
\newblock {\em J. ACM}, 41(4):764--790, July 1994.

\bibitem{ss-ampr-90}
J.~T. Schwartz and M.~Sharir.
\newblock Algorithmic motion planning in robotics.
\newblock In J.~van Leeuwen, editor, {\em Algorithms and Complexity}, volume~A
  of {\em Handbook of Theoretical Computer Science}, pages 391--430. Elsevier,
  Amsterdam, 1990.

\bibitem{Sut89}
W.~R.~S.~Sutherland
\newblock The offsets of a convex polygon.
\newblock {\em Methods of Operation Research}, 62:33-41, 1989.

\bibitem{Swo91}
E.~W.~Swokowski.
\newblock {\em Calculus with Analytic Geometry}.
\newblock PWS-Kent, Boston, 1991.

\bibitem{w-mpav-88}
G.~Wilfong.
\newblock Motion planning for an autonomous vehicle.
\newblock In {\em Proc. IEEE Internat. Conf. Robot. Autom.}, pages 529--533,
  1988.

\end{thebibliography}
\end{document}